\documentclass[12pt]{article}

\usepackage[a4paper, margin=1in]{geometry}
\usepackage{titling}           
\usepackage{tocloft}           
\usepackage{titlesec}          
\usepackage{amsmath, amssymb}  
\usepackage{lmodern}           
\usepackage{setspace}          
\usepackage{hyperref}          
\usepackage{graphicx}
\usepackage{indentfirst}
\usepackage{caption}
\usepackage{blkarray} 
\usepackage{cite}
\usepackage{abstract} 
\usepackage{pdfpages}

\renewenvironment{abstract}{
  \begin{center}
    \normalfont\normalsize\bfseries  
    \abstractname\vspace{-.5em}\vspace{0pt}
  \end{center}
  \list{}{
    \setlength{\leftmargin}{0cm}
    \setlength{\rightmargin}{0cm}
  }
  \item\relax\bfseries  
}{\endlist}

\title{\LARGE\bfseries Strategy evolution on networks \\ under payoff uncertainty and risk preference}
\date{}

\cftsetindents{section}{0em}{4em}
\cftsetindents{subsection}{2em}{4em}

\titleformat{\section}{\normalfont\large\bfseries}{\thesection}{1em}{}
\titleformat{\subsection}{\normalfont\large}{\thesubsection}{1em}{}

\usepackage{authblk} 

\author[1,2]{Jiapeng Yu}
\author[3,4]{Anzhi Sheng}
\author[1,2]{Long Wang}

\affil[1]{Center for Systems and Control, College of Engineering, Peking University, Beijing, China.}
\affil[2]{Center for Multi-Agent Research, Institute for Artificial Intelligence, Peking University, Beijing, China.}
\affil[3]{Department of Decision and Control Systems, KTH Royal Institute of Technology, Stockholm, Sweden.}
\affil[4]{Department of Mathematics, KTH Royal Institute of Technology, Stockholm, Sweden.}

\begin{document}

\maketitle

\begin{abstract}
Cooperation is a key driver of human social progress. 
Studies of the evolution of cooperation typically assume a deterministic outcome for social interactions.
But in real-world social interactions, interaction outcomes are often subject to stochastic perturbations arising from open environments. 
Individuals may show different attitudes towards such uncertainty, some are risk-seeking, while others tend to be risk-averse.
Here we investigate how risk preference towards uncertain payoffs affects the evolution of cooperation on social networks, where uncertainty originates from random punishment of defectors initiated by cooperators. 
We provide an analytical treatment of how the distribution of risk preference among individuals alters the threshold required for cooperation. 
We find that, at the population level, risk-averse behavior promotes or even rescues cooperation. 
At the node level, variation in risk preference has a significant impact when it occurs on nodes with high degree centrality. When nodes have the same degree centrality, the nodes with lower betweenness centrality exhibit a stronger effect on strategy evolution.
Our analysis reveals how risk preference, together with spatial structure, jointly shapes and potentially reverses the evolutionary dynamics of cooperation.

\end{abstract}

\section{Introduction}
Cooperation is key to the development and prosperity of social systems, ranging from animal groups to human society \cite{Perc2017, West2007, Stander1992,Packer1988,Wuchty2007}.
But cooperation remains vulnerable to invasion by defection, as individuals often prioritize maximizing their own interests over helping others.
Therefore, how cooperation emerges and prevails over defection has long been seen as a fundamental and practically important question \cite{Perc2017}. 
Evolutionary game theory provides a powerful framework for modeling the evolution of strategies in complex systems, where evolutionary dynamics can be decomposed into two phases: individual interactions and strategy imitation.
Using this theory, many studies have revealed that population structure is one of the most important mechanisms to promote the emergence of cooperation in real-world social interactions \cite{Szabo2007,Nowak1992,Nowak2006,Allen2017,Nowak2004,Taylor2007,Tarnita2009,Su2022,Sheng2023,Ohtsuki2006,Lieberman2005,Fu2010,Santos2008,Santos2005,wang2023stability,wang2024deterministic,chen2008promotion}.
This theoretical insight has been further supported by several experimental studies \cite{rand2011dynamic,rand2013human}.

The literature on network reciprocity typically assumes that interactions occur in deterministic and static environments, in which outcomes, such as individuals' payoffs, are determined by the current population state and structure. 
This assumption overlooks the pervasive uncertainty inherent in real-world social interactions, arising from environmental noise \cite{assaf2013cooperation, van2018uncertainty}, fluctuations in individual behavior \cite{rand2015thought}, and heterogeneous personalities. 
As a result, payoffs can be stochastic even when the population state and structure are fixed.
In this context, we highlight a notable work \cite{Wang2024} that incorporates individual risk preferences and uses utility theory to characterize how individuals, with different risk preferences, respond to payoff uncertainty and make decisions.
One limitation of this work is that it assumes an idealized, infinitely large well-mixed population, rather than a more realistic structured social network.
Another limitation is that payoff uncertainty is modeled as an abstract Gaussian random variable, rather than arising from real social behavior.

In this work, we model payoff uncertainty by punishment imposed on defectors by cooperators.
In the real world, punishment is often implemented stochastically, as it might be limited by imperfect information, communication delays, and/or individuals’ willingness to act \cite{boyd2003evolution, henrich2006costly}.
Unlike traditional studies that investigate the role of punishment on cooperation \cite{boyd2003evolution}, where punishment is treated as an additional strategy alongside cooperation and defection, here punishment is modeled as a countermeasure adopted by cooperators against defectors. As a result, a defector can obtain a relatively high payoff in the absence of punishment, but a low payoff when punishment is imposed. This risk is public information to all individuals, and this is the place where risk preferences come into play. 
A risk-neutral individual evaluates the utility as the exact expected value of the stochastic payoff, whereas risk-averse or risk-seeking individuals place more weight on outcomes with or without punishment. 

Here, we develop a theoretical framework of evolutionary game dynamics that incorporates population structures and payoff uncertainty modeled by punishment. The heterogeneity in individuals' attitudes towards risk and uncertainty is modeled by their different risk preferences. We derive the threshold for cooperation to evolve on arbitrary network structures and risk preference distributions.
Building on these theoretical results, we investigate how the distribution of risk preferences, coupled with the distribution of node degrees, affects the evolution of cooperation at both the population and node levels across stylized and a broad range of random networks.
We find that introducing payoff uncertainty and risk preferences can promote cooperation, and even rescue cooperation in situations where cooperation would otherwise fail to evolve under the traditional model. Our analysis reveals how population structure, together with risk preferences, jointly shape the direction of strategy evolution.

\section{Model}
\subsection{Interactions with probabilistic punishment}

\begin{figure}[htbp]
    \centering
    \includegraphics[width=\textwidth]{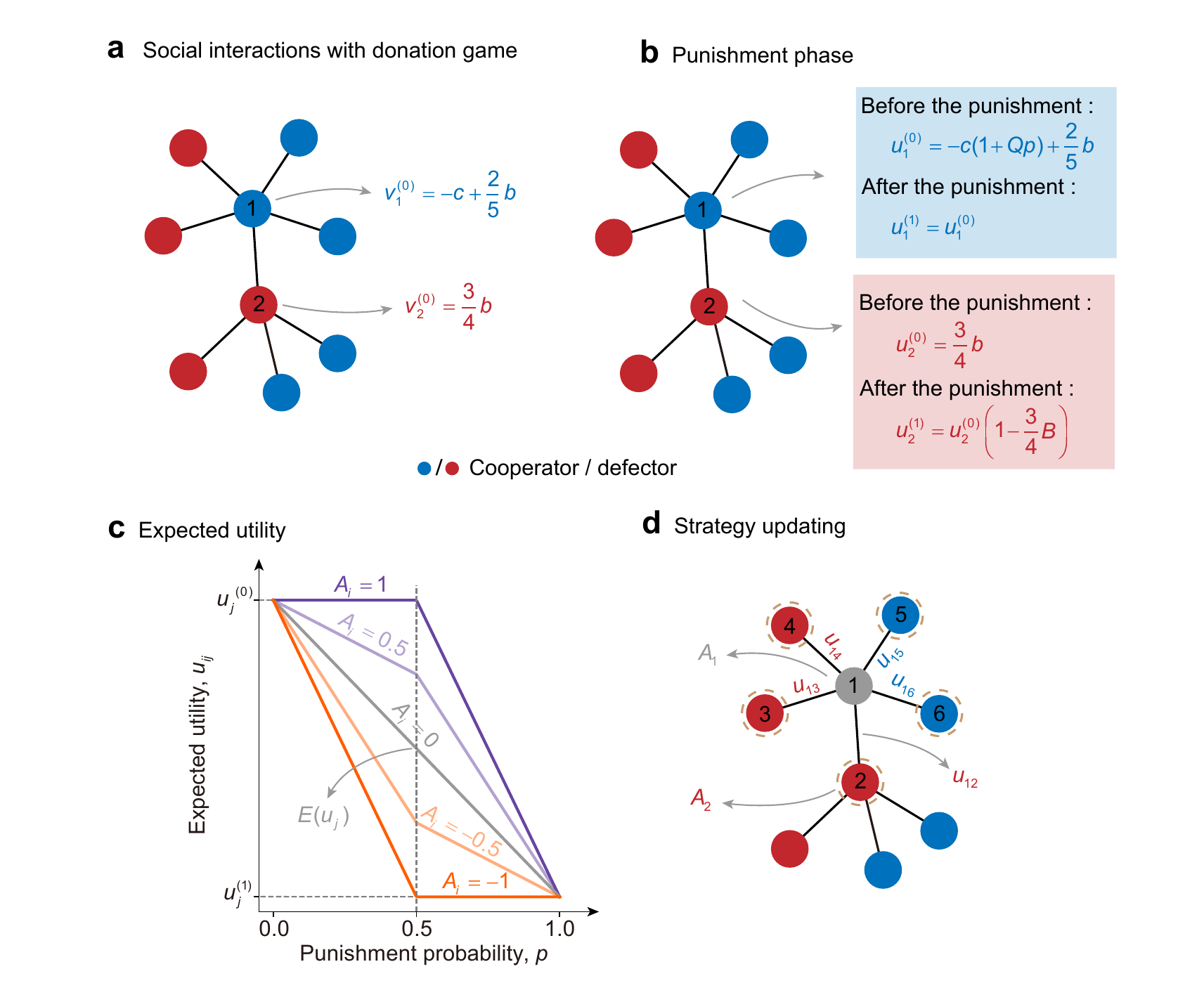}
    \captionsetup{labelformat=empty} 
    \caption{\textbf{Fig.~1|Evolutionary dynamics with risk preference and punishment.} 
    \textbf{a,} A population is represented by a graph, where nodes represent individuals and edges represent interactions. At each time step, each individual $i$ adopts either cooperation (blue circles) or defection (red circles) and obtains an initial payoff $v_{i}^{(0)}$ by playing the donation game with their neighbors.
    \textbf{b,} Subsequently, each cooperator pays a fixed cost to punish each defector by discounting their payoff. This punishment is stochastic with probability $p$.
    Therefore, the final payoff of a cooperator is deterministic, while the final payoff of a defector is stochastic.
    \textbf{c,} Individuals hold different attitudes towards the uncertainty in payoffs, which is captured by risk preference $A_i\in[-1,1]$:  $A_i=0$ corresponds to a neutral baseline, where the payoff of $j$ from individual $i$'s perspective, $u_{ij}$, is a linear combination of the payoffs before and after punishment weighted by $p$ (grey line),
    while $A_i > 0$ ($A_i<0$) corresponds to risk-seeking (risk-averse) attitude where $u_{ij}$ is a concave (convex) function with respect to $p$. 
    \textbf{d,} Individual $1$ (grey circle) is randomly chosen to update their strategy by imitating one of their neighbors $j$ (brown dashed circle), with probability proportional to $j$'s fitness $F_{1j}=\exp(\delta u_{1j})$.}
    \label{fig:1}
\end{figure}

We consider a population of $N$ individuals, labeled as $\mathcal{N} = \{1,2,\dots,N\}$.
The population is represented by a graph, where nodes represents individuals and edges denote pairwise interactions between them. Let $ W = \left(w_{ij}\right)_{i,j=1}^N$ be the adjacency matrix of this graph, where the entry \( w_{ij} =1 \) if there exists a link between individual $i$ and $j$, and $w_{ij}=0$ otherwise (see Fig.~\ref{fig:1}a). 

In each round, every individual chooses either cooperation (C) or defection (D) to interact with each of their neighbor. The strategy of an individual $i$ is denoted by $x_{i}$, where $x_{i}=1$ represents cooperation and $x_{i}=0$ represents defection.  
Individuals first engage in the donation game with their neighbors: a cooperator pays a cost $c$ to provide a benefit $b$ for neighbors, and a defector pays no cost and provides no benefit.  
As a result, the payoff of individual $i$ is 
\begin{equation}
     v_{i}^{(0)}=-cx_{i}+b\sum_{j=1}^{N}p_{ij}^{(1)}x_{j}.
\end{equation}

After all interactions are completed, each cooperator pays an additional cost $\alpha$ to punish each defector with probability $p$, reducing their payoff to a fraction $\beta_i$ of its original value.
Specifically, let \(u_{i}^{(0)}\) and \(u_{i}^{(1)}\) denote the payoff of individual \(i\) before and after the punishment, respectively. We have
\begin{equation}
    \begin{aligned}
        u_i^{(0)} &= v_i^{(0)} - \alpha x_i,\\
        u_i^{(1)} &= x_{i}u_{i}^{(0)}+(1-x_{i})\beta_i u_i^{(0)}.
    \end{aligned}
\end{equation}
Then the final payoff of \(i\) takes the form of
\begin{equation}
    u_i = u_i^{(1)} X + u_i^{(0)} \left(1 -X \right),
    \label{eq:random_payoff}
\end{equation}
where $X$ follows a Bernoulli distribution with parameter $p$.
If $i$ is a cooperator, $u_i$ is a deterministic variable as \(u_{i}^{(0)} = u_{i}^{(1)}\).
Conversely, if $i$ is a defector, $u_i$ becomes a random variable (see Fig.~\ref{fig:1}b). 

In this work, the additional cost $\alpha$ is defined as $\alpha = cpQ$, meaning that cooperators must pay a greater cost to increase the probability of punishment.
The strength of punishment $\beta_i$ depends on the number of cooperative neighbors surrounding $i$, defined as $\beta_{i} = 1 - \sum_{j=1}^N p_{ij}^{(1)} x_j B$, which indicates that the strength is stronger if there are more cooperative neighbors.
Here $Q \in [0,1]$ and $B\in [0, 1]$ are free parameters.

\subsection{Strategy updating based on individual risk preference}

Following the death-birth updating rule \cite{zukewich2013consolidating} at the end of each round, a random individual $i$ is selected to update their strategy by copying one neighbor, denoted by $j$. The imitation probability is positively correlated with $j$'s payoff $u_j$.
However, as $u_j$ can be a random variable, individuals may exhibit different attitudes towards the resulting uncertainty. 
Here we introduce the risk preference \(A_{i} \in [-1,1]\) to quantify $i$'s attitude towards uncertain payoffs.
The neutral baseline $A_i=0$ means that the utility of $j$'s strategy from $i$'s perspective, denoted by $u_{ij}$, is exactly the expectation of $u_j$.
The positive case \(A_{i} > 0\) corresponds to risk-seeking behavior where the risky payoff $u_j^{(0)}$ is over-weighted, compared to the neutral baseline. 
Conversely, the negative case corresponds to risk-averse behavior where the conservative payoff $u_j^{(1)}$ is over-weighted.

For analytical tractability, we adopt a piecewise linear form for $u_{ij}$ (see Fig.~\ref{fig:1}c) in the main text, which can serve as approximations to general smooth form of utility \cite{murthy1998stochastic}:
\begin{equation}
    u_{ij}=\begin{cases}
        u_{j}^{(0)}-p(1-A_{i})\left ( u_{j}^{(0)}-u_{j}^{(1)} \right ) &  p\le 0.5,\\
        u_{j}^{(1)}+(1-p)(A_{i}+1)\left ( u_{j}^{(0)}-u_{j}^{(1)} \right )   &  p\ge 0.5.
    \end{cases}
    \label{eq:uij}
\end{equation}
The above function reveals the features of expected utility with risk preference: when $p = 0$ or $p = 1$, there is no uncertainty in the payoffs, so the expected utility coincides with the payoff value; when $p = 0.5$, the uncertainty is maximized, and the difference between the expected utility and the averaged weighted payoff is also maximized.
We also analyze the evolution of cooperation based on a smooth form for $u_{ij}$ in SI (Supplementary Fig.~1).
The fitness of $j$ from $i$'s perspective is $F_{ij}=\exp (\delta u_{ij})$, where $\delta \ge 0$ is the intensity of selection, measuring the influence of expected utility on strategy updating \cite{traulsen2008analytical}. The probability that individual $i$ copies $j$'s strategy is
\begin{equation}
    p_{j \to i}=\frac{1}{N}\frac{w_{ji}F_{ji}}{ {\textstyle \sum_{l=1}^{N}} w_{jl}F_{jl}}.
\end{equation}
As shown in Fig.~\ref{fig:1}d, when a random individual $1$ is selected, she evaluates the fitness of her neighbors based on her risk preference $A_1$, and adopts one of their strategies with a probability proportional to their respective fitness. 
In the subsequent analysis, we focus on the regime of weak selection ($\delta \ll 1$).

\section{Results}
\subsection{General rule for the evolution of cooperation with risk preference}

In the absence of mutation, the population will eventually reach one of two homogeneous states: the all-cooperator state or the all-defector state. Let $\rho_{C}$ denote the fixation probability of cooperation: the probability that a single cooperator, randomly introduced into a population of defectors, eventually dominates the population. We say cooperation is favored by natural selection if the fixation probability exceeds that under neutrality ($\delta=0$), that is, $\rho_{C} > 1/N$. 

We prove that cooperation is favored if
\begin{equation}
    \left(f_{b}^{(1)}+f_{b}^{(2)}\right)b-
    \left(f_{c}^{(1)}+f_{c}^{(2)} \right)c>0.
    \label{eq:condition}
\end{equation}
Here, \(f_{b}^{(n)}\) and \(f_{c}^{(n)}\) ($n=1,2$) denote the coefficients associated with the benefit and cost induced by cooperative behavior. 
Specifically, $f_b^{(1)}$ and $f_c^{(1)}$ correspond to the benefit and cost in the traditional model (without punishment and risk preferences), whereas \(f_{b}^{(2)}\) and \(f_{c}^{(2)}\) denote the additional benefit and cost that arise when payoff uncertainty and risk preference are taken into account (see Methods for analytical expressions). These additional terms appear because cooperators need to pay an extra cost to punish defectors, thereby increasing their cost.
Meanwhile, defectors face a risk of punishment, which reduces their net payoffs and increases the relative benefit of cooperators.
In conclusion, Eq.~\ref{eq:condition} indicates that cooperation is favored by natural selection when the total benefit of cooperators exceeds their total cost. 

The critical benefit-to-cost ratio required for cooperation to evolve is given by
\begin{equation}
    \left(\frac{b}{c} \right)^* = \frac{f_{c}^{(1)}+f_{c}^{(2)}}{f_{b}^{(1)}+f_{b}^{(2)}}.
\end{equation}
A positive and smaller critical benefit-to-cost ratio implies stronger promotion of cooperation, since a relatively lower benefit-to-cost ratio $b/c$ is required to make cooperation selectively advantageous.  
Then, the central question of this study is, given a population structure, how do variations in the distribution of risk preference increase or decrease the critical benefit-to-cost ratio $(b/c)^*$, relative to that without risk preference and punishment (i.e. the traditional model).

\subsection{Stylized population structures}
\begin{figure}[htbp]
    \centering
    \includegraphics[width=\textwidth]{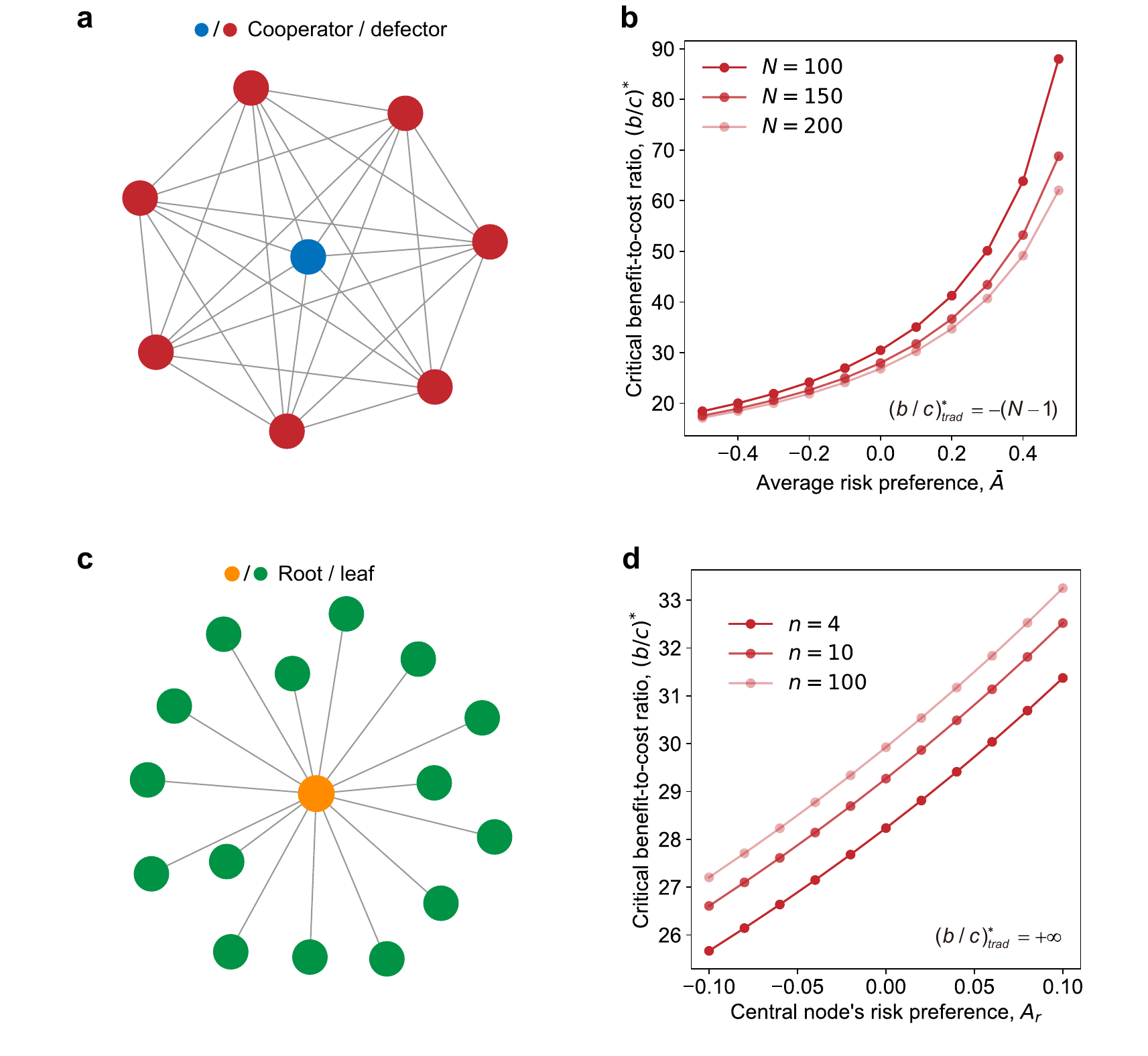}
    \captionsetup{labelformat=empty} 
    \caption{\textbf{Fig. 2|Risk-averse behavior promotes cooperation in both star and well-mixed populations.} 
    \textbf{a,} We consider a classical homogeneous population -- a well-mixed population of $N$ individuals, where each individual interacts with all others. The evolutionary process begins with a single cooperator in a fully defective population.
    \textbf{b,} In well-mixed populations, the effect of the distribution of risk preferences on cooperation is captured by its first moment $\bar{A}$ (see Eq.~\ref{eq:wellmixed}).
    We calculate the critical benefit-to-cost ratio $(b/c)^*$ as a function of the population size $N$ and average risk preference  $\bar{A}$.
    The critical ratio increases monotonically with $\bar{A}$ for a fixed $N$, meaning that a more risk-seeking population hinders cooperation.
    But compared to the critical ratio under the traditional setting, incorporating risk preference and payoff uncertainty can rescue cooperation by shifting the critical ratio from a negative value ($(b/c)^*_{trad}=-(N-1)$) to a positive one.
    \textbf{c,} We also consider an extremely heterogeneous population, represented by a star network, which consists of a central root node (orange circle) and $n$ leaf nodes (green circles). Then, we have $N=n+1$.
    \textbf{d,} On star networks, the critical ratio $(b/c)^*$ only depends on the risk preference of the root node $A_r$ (see Eq.~\ref{eq:star}). Similar to \textbf{a}, a higher value of $A_r$ induces a higher critical ratio, which hinders cooperation. But overall, the introduction of risk preference and punishment promotes cooperation compared to the traditional model, as the critical ratio changes from an infinite value $(b/c)^*_{trad}=\infty$, where cooperation can never evolve, to a finite positive value. 
    Parameters: $Q=0$, $p=0.5$ and $B=0.5$ in \textbf{b}, and $B=0.2$, $Q=0$, and $p=0.5$ in \textbf{d}.
    }
    \label{fig:2}
\end{figure}

 \textbf{Well-mixed populations.}
We first consider a well-mixed population, where each individual is connected to all other individuals (see Fig.~\ref{fig:2}{a}). We first derive a closed-form expression for the critical benefit-to-cost ratio,
\begin{equation}
\left(\frac{b}{c}\right)^{*}=
\begin{cases}
\dfrac{6(1+Qp)(N-1)^{2}}{BN(N+1)p\left(1-\bar{A}\right)-6(N-1)}, & p \leq 0.5, \\[1.0em]
\dfrac{6(1+Qp)(N-1)^{2}}{BN(N+1)p\left[1-(1-p)\left(1+\bar{A}\right)\right]-6(N-1)}, & p \geq 0.5.
\end{cases}
\label{eq:wellmixed}
\end{equation}
When the number of nodes $N$ approaches infinity and $p$ and $Q$ take arbitrary values within their respective domains, the critical ratio lies within the interval $\{-\infty\} \cup \left[ 6/B, +\infty \right)$ (see Supplementary Section~3.1 for detailed derivations).

By Eq.~\ref{eq:wellmixed}, we find that for a fixed evolutionary setting (fixed parameters $p$, $B$, and $Q$), the critical benefit-to-cost ratio $(b/c)^*$ is directly determined by the average risk preference of the population, $\bar{A} = \sum_{i=1}^N A_i/N$. In other words, in a completely homogeneous population structure, the evolution of cooperation depends only on the first moment of the distribution of risk preference. Furthermore, the critical ratio increases monotonically with $\bar{A}$ (see Fig.~\ref{fig:2}{b}) for any given penalty severity $B$, meaning that a more risk-averse population (on average) is more favorable to the evolution of cooperation. 
The intuition behind this phenomenon is that risk-averse individuals are more likely to resist the temptation of potentially high rewards obtained by defection, and therefore maintain the stability of cooperation. 

In addition, when the population is not entirely risk-seeking (i.e.~$\bar{A} \neq 1$), the critical ratio can be positive in large populations, as long as the possibility of punishment exists ($p>0$).
Even when the cost of punishment is substantial (large $Q$), and the penalty severity to defectors is minimal (small $B$), the introduction of potential punishment, together with heterogeneous risk preferences, rescues cooperation in well-mixed populations by reversing the dynamics from favoring spiteful behavior (negative $(b/c)^*$ under the traditional setup) \cite{fehr2002altruistic,boyd2003evolution,sigmund2010social} to prosocial behavior (positive $(b/c)^*$).

 \textbf{Star networks.}
We have investigated the effect of risk preference on cooperation in completely homogeneous, well-mixed population structures, where only the first moment of the risk preference distribution is at work.
We now move to analyze whether and how the individual heterogeneity in
risk preference affects cooperation in heterogeneous population structures.
We begin with the extremely heterogeneous network structure -- a star graph, which consists of one central root node and $n$ leaf nodes (see Fig.~\ref{fig:2}{c}).
In this case, the critical benefit-to-cost ratio is given by  
\begin{equation}
    \left ( \frac{b}{c} \right ) ^{*}=\begin{cases}
        \dfrac{12n(1+Qp)}{(4n+1)Bp(1-A_{r})},   & p\le0.5,\\[6pt]
        \dfrac{12n(1+Qp)}{(4n+1)B\left [ 1-(1-p)(1+A_{r}) \right ] },   & p>0.5. \\
    \end{cases}
    \label{eq:star}
\end{equation}
where $A_{r}$ denotes the risk preference of the root node. 
When the number of nodes $N$ approaches infinity and $p$ and $Q$ and $B$ take arbitrary values within their respective domains, the critical ratio $(b/c)^{*}$ lies within the range $[3, +\infty)$ (see Supplementary Section 3.2 for detailed derivations).

Equation \ref{eq:star} shows that the critical ratio $(b/c)^*$ increases monotonically with the risk preference of the root node $A_r$ (see Fig.~\ref{fig:2}{d}).
This implies that the evolution of cooperation depends solely on the attitude of the root node towards risk and uncertainty. Specifically,
a more risk-averse root node creates more favorable conditions for cooperation. 

This finding, in general, agrees with the statement: a lower risk preference parameter promotes cooperation, but deviates from that observed in well-mixed populations in detail. For instance, when the average risk preference $\bar{A}$ becomes lower, the critical ratio $(b/c)^*$ in well-mixed populations decreases, and therefore cooperation is easier to evolve. But in the star graph, the risk preference of the root node can remain unchanged or even increase, even when $\bar{A}$ decreases. As a result, $(b/c)^*$ is unchanged or becomes higher. 
In other words, the way the risk preference distribution affects cooperation is highly coupled with population structures, which highlights the effect of individual heterogeneity, in both spatial location and risk preference, on evolutionary outcomes.

There is one important finding in the star graph, that coincides with that in well-mixed populations: when the root node is not entirely risk-seeking (i.e. $A_r\neq 1$ or $\bar{A}\neq 1$), the introduction of punishment and risk preference rescues cooperation by lowering $(b/c)^*$ from infinity (as seen under the tradition setup, where cooperation is never favored for any benefit-to-cost ratio $b/c$) to a finite positive value (where cooperation can be favored if $b/c$ is higher than $(b/c)^*$) (see Fig.~\ref{fig:2}{d}).

\subsection{Impact of highly connected nodes}
\label{section:nodedegree}

\begin{figure}[htbp]
    \centering
    \includegraphics[width=\textwidth]{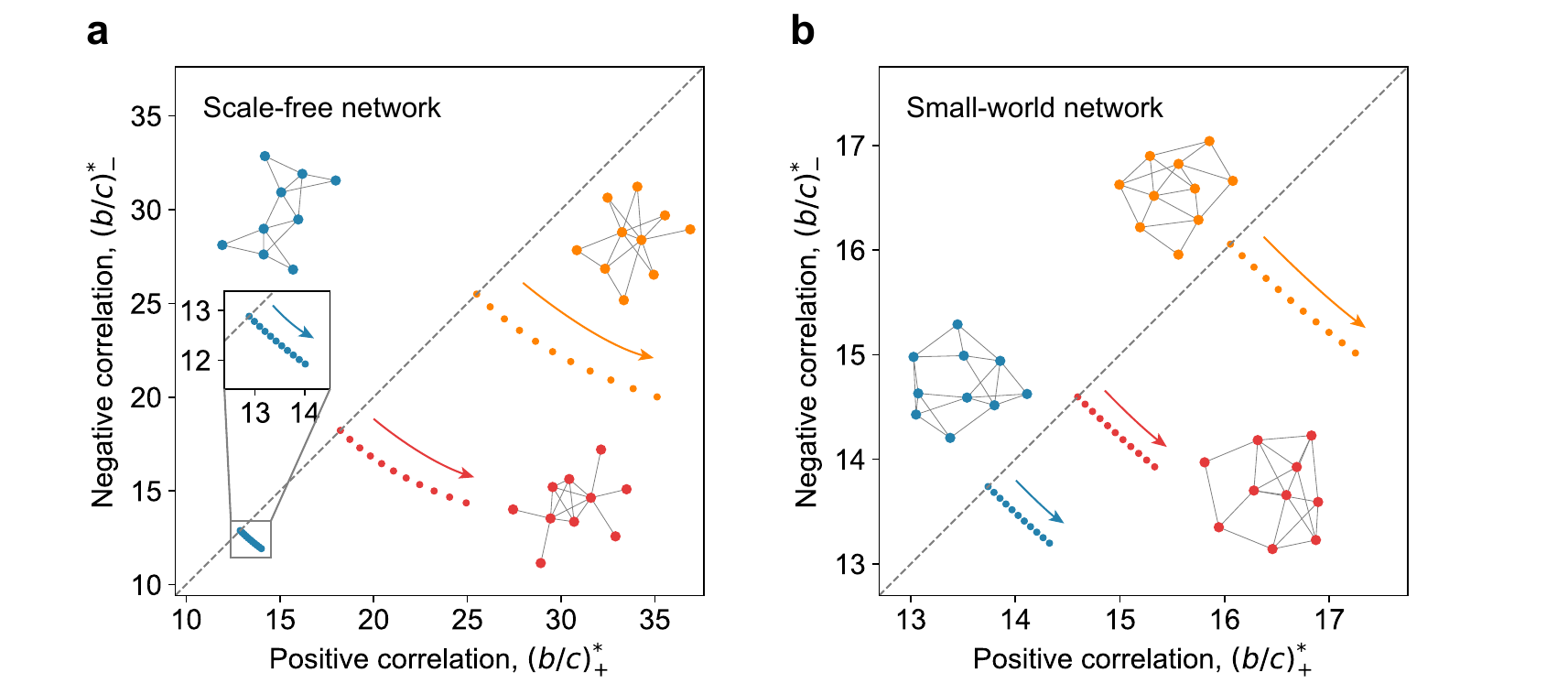}
    \captionsetup{labelformat=empty} 
     \caption{\textbf{Fig.~3|Risk preference of high-degree nodes has a stronger impact on the evolution of cooperation.} 
     We analyze how the distributions of node degrees and risk preferences collectively influence the evolution of cooperation. 
    We consider six random networks: three are scale-free networks (\textbf{a}) \cite{barabasi1999scalefree}, and the remaining three are small-world networks (\textbf{b}) \cite{watts1998smallworld}.
    We fix the mean of the risk preference distribution at neutrality (i.e. $\bar{A}=0$) and change its variance.
    As the variance increases (the direction of the arrows), the critical ratio increases under positive correlation between the two distributions, but decreases under negative correlation.
    This result indicates that both the cooperation-hindering and cooperation-promoting effects are more strongly governed by the risk preferences of high-degree nodes. Parameters: $B=0.2$, $p=0.5$, and $Q=0$ in both \textbf{a} and \textbf{b}.}
    \label{fig:3}
\end{figure}

From homogeneous well-mixed populations to heterogeneous star networks, we find that population structures affect how the distribution of risk preferences functions in the evolution of cooperation. 
Nodes with higher degrees have a stronger effect on the critical ratio. 
To verify this observation, we change the variance of the risk preference distribution (while keeping the mean neutral, $\bar{A}=0$), and analyze how the critical ratio changes on the six random graphs, three of which are scale-free networks (Fig.~\ref{fig:3}{a}) \cite{barabasi1999scalefree}, and the remaining three are small-world networks (Fig.~\ref{fig:3}{b}) \cite{watts1998smallworld}. 
Specifically, we consider two cases: 1. the distributions of risk preferences and node degrees are positively correlated (i.e. high-degree nodes are more risk-seeking); 2. they are negatively correlated (i.e. high-degree nodes are more risk-averse).
The corresponding critical ratios are denoted $(b/c)_{+}^{*}$ and $(b/c)_{-}^{*}$, respectively.

In the positive correlation case, increasing the variance of risk preferences raises the critical ratio $(b/c)_{+}^{*}$, while in the negative correlation case, it lowers the critical ratio $(b/c)_{-}^{*}$. This phenomenon further verifies our conclusions from the stylized networks: 1. the risk preferences of high-degree nodes play a dominant role in shaping the evolution of cooperation; and 2. risk-averse behavior can promote cooperation. 
The intuition behind these results is that high-degree nodes can transmit their strategies to more individuals, and therefore have a stronger effect on evolutionary dynamics. When these high-degree nodes are risk-averse, they are more resilient to invasion by defectors, while being more likely to transmit cooperation to their neighbors. As a result, the population is more likely to reach the fully cooperative state (see Supplementary Fig.~2).

We further validate our findings within more random graphs (see Supplementary Fig.~3) and by an alternative approach -- tracking variations in the critical ratios as the risk preference of a single node is varied while those of the remaining nodes are fixed (see Supplementary Fig.~4). The results consistently support the same conclusion: nodes with higher degrees exert a stronger influence on the critical ratio.

\begin{figure}[htbp]
    \centering
    \includegraphics[width=\textwidth]{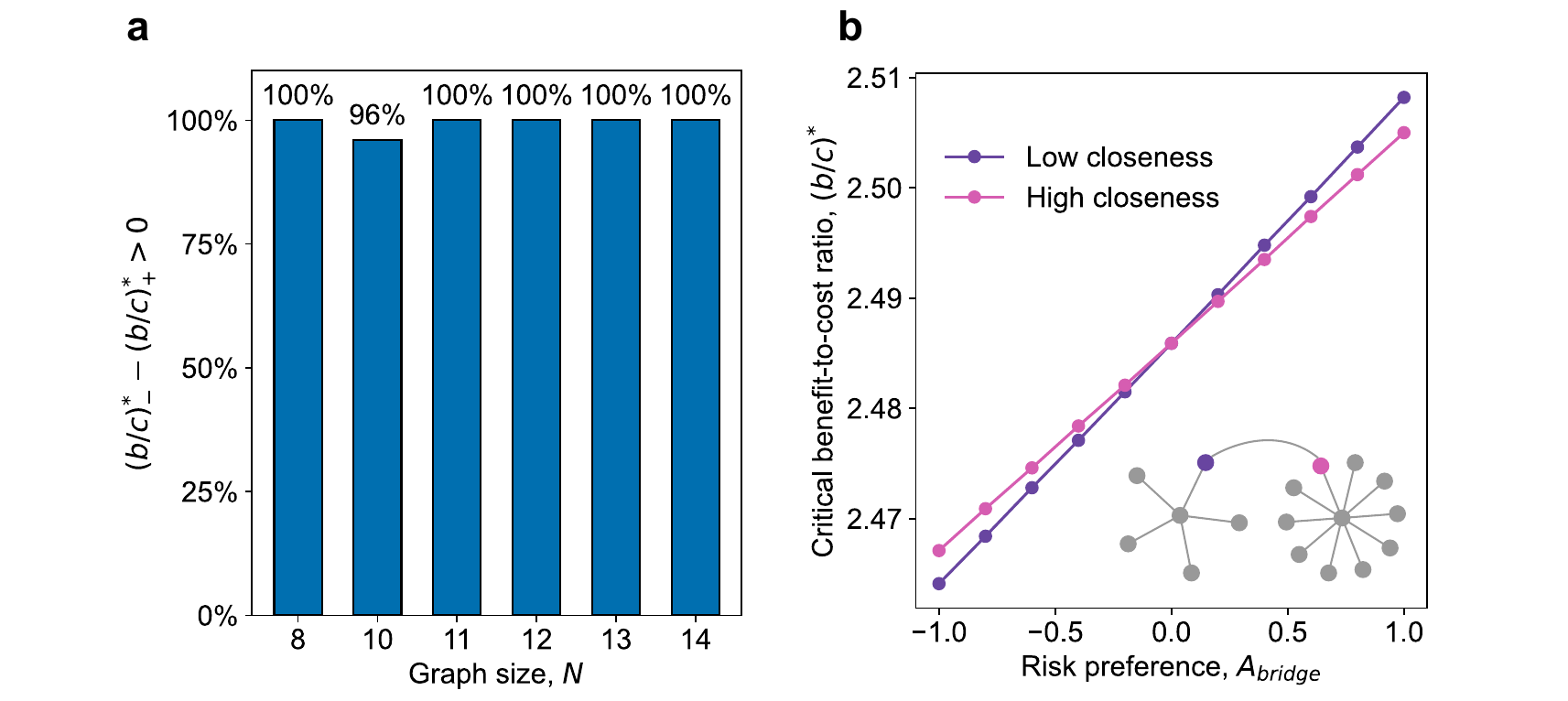}
    \captionsetup{labelformat=empty} 
    \caption{\textbf{Fig.~4|Nodes with a low betweenness centrality have a stronger impact on the evolution of cooperation.} 
    \textbf{a,} We generate $600$ random regular graphs of six different sizes, with 100 graphs for each size.
    For each graph, risk preferences are assigned either positively or negatively proportional to node degrees, but the average risk preference is kept neutral. 
    The results show that the critical ratios in the negatively related case ($(b/c)_{-}^{*}$) are lower than those in the positively related case ($(b/c)_{+}^{*}$), indicating that risk preferences of low-betweenness nodes have a stronger effect on cooperation. 
    \textbf{b,} Two star components are connected via their corresponding leaf nodes, which are referred to as bridge nodes (purple circles). 
    We keep the risk preferences of all other nodes neutral while varying those of the two bridge nodes, denoted $A_{bridge}$. The results show that the curve for the node with lower betweenness centrality (i.e. the left bridge node) exhibits steeper slopes with respect to $A_{bridge}$, further supporting our previous conclusion.
    Parameters: $B=0.2$, $p=0.5$, and $Q=0$ in \textbf{a}, and $B=0.5$, $p=0.5$, and $Q=0$ in \textbf{b}.}
    \label{fig:4}
\end{figure}

\subsection{Impact of betweenness centrality}

We have shown that a node's degree (a local topological property) determines the extent to which its risk preference affects the evolution of cooperation. 
We next investigate how a more global node property affects cooperation. 
Specifically, for a given network, we group nodes by degree and analyze the correlation between their betweenness centrality \cite{freeman1977set,freeman1978centrality,brandes2001faster} and their inflence on cooperation. 
Betweenness centrality quantifies how frequently a node lies on the shortest paths between pairs of nodes (see Methods for details).
Therefore, nodes with high betweenness centrality act as bridges connecting different parts of the network and may govern strategy evolution in the network.

We first consider a broad range of random regular graphs, where each node has an identical degree (Fig.~\ref{fig:4}{a}).
Similar to section \ref{section:nodedegree}, we keep the mean risk preference neutral ($\bar{A} = 0$) and assign the distribution of risk preferences in two ways: one positively proportional to the betweenness centrality of nodes, and the other negatively proportional to it.
Our result shows that the critical ratio in the first case, denoted $(b/c)_{+}^{*}$, is generally lower than that in the second case, denoted $(b/c)_{-}^{*}$ (see Fig.~\ref{fig:4}{a}), which means that nodes with lower betweenness centrality has a stronger effect on the evolution.
We further verify this conclusion on a bi-star network (Fig.~\ref{fig:4}{b}), where two star components of different sizes are connected by their respective leaf nodes.
We call these two nodes bridge nodes.
We find that changes in the critical ratio $(b/c)^*$ are more sensitive to changes in the risk preference of the bridge nodes in the smaller star component (i.e. that with a lower betweenness centrality).
This finding further supports our previous conclusion. 

\begin{figure}[htbp]
    \centering
    \includegraphics[width=\textwidth]{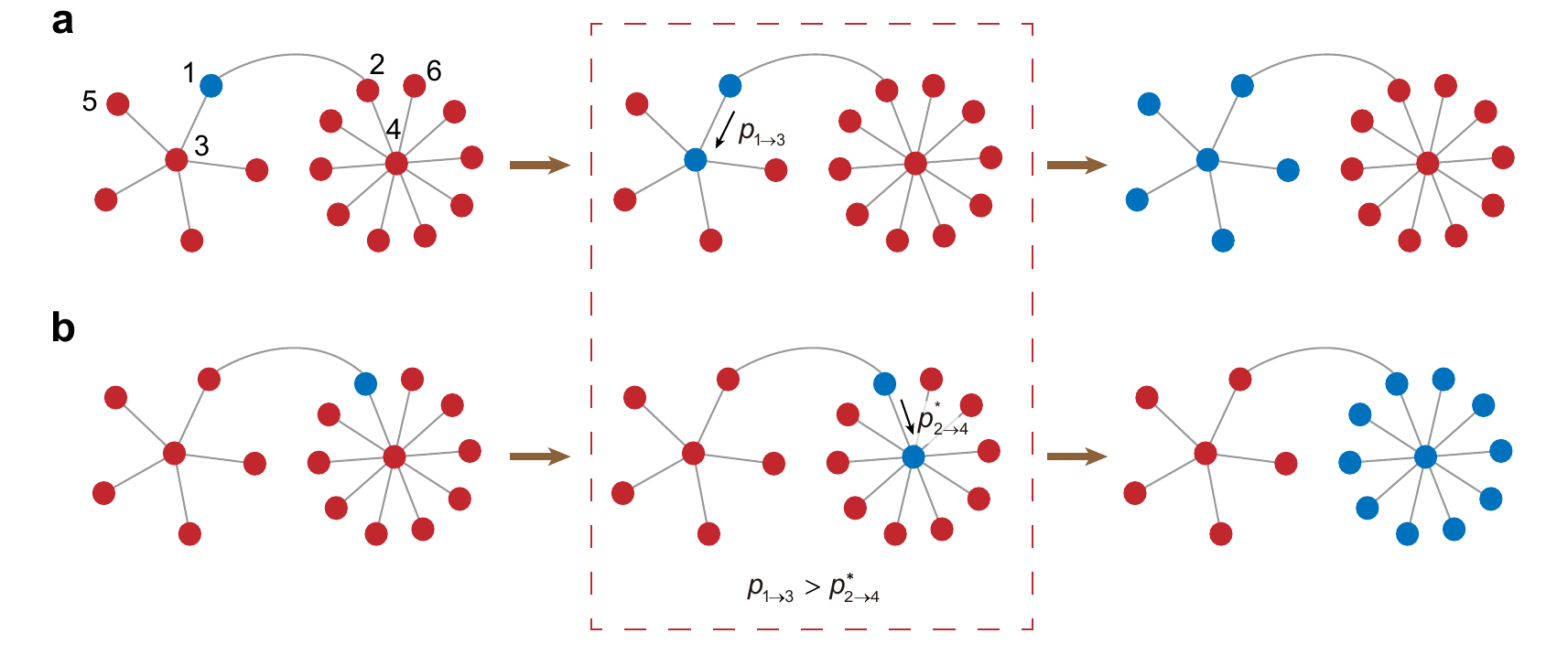}
    \captionsetup{labelformat=empty} 
    \caption{\textbf{Fig.~5|Intuition for how nodes with lower betweenness centrality dominate the evolutionary direction.} 
    We consider two evolutionary trajectories on a bi-star network composed of full defectors: one in which a cooperator is introduced at the bridge node with lower betweenness centrality (node 1, \textbf{a}); and the other in which it is introduced at the bridge node with higher betweenness centrality (node 2, \textbf{b}).
    We find that the probability of strategy transmission from node 1 to node 3 (denoted $p_{1\to3}$) is higher than that from node 2 to node 4 (denoted $p^{*}_{2\to4}$), which means nodes with lower betweenness centrality are more likely to transmit their strategy to the hub node in their component. 
    This property facilitates the spread of cooperation throughout the component and ultimately enables it to occupy the entire population.
    In other words, nodes with lower betweenness centrality are more likely to transmit their strategies throughout the population, and therefore have a stronger effect on the evolutionary direction.
    }
    \label{fig:5}
\end{figure}

To illustrate why nodes with lower betweenness centrality have a stronger effect on strategy evolution, we compare two evolutionary trajectories in which an initial cooperator is located at each of the two bridge nodes of an otherwise fully defective bi-star network (Fig.~\ref{fig:5}).
We find that the probability of transmitting cooperation from the bridge node to the nearest root node is higher when the bridge node belongs to the smaller star component (i.e. has lower betweenness centrality).
As a result, cooperation more readily spreads throughout the smaller-sized star component, which in turn fosters its spread to the other star component. 
In other words, nodes with lower betweenness centrality have a greater capability to propagate their strategy throughout the whole population.
As individuals are more (or less) likely to sustain cooperation when their risk preference is lower (or higher), variations in the threshold for cooperation are more significant as the risk preference of the bridge node with lower betweenness centrality varies (Fig.~\ref{fig:4}b). 

\section{Discussion}

Real-world interactions are often affected by environmental disturbances, communication noise, and other sources of uncertainty. 
Therefore, their outcomes are more appropriately modeled as random variables rather than deterministic values, and individuals' attitudes towards such uncertainty can substantially affect the direction of strategy evolution.
An existing theoretical study has revealed that individuals' risk preferences play a key role in evolutionary dynamics of cooperation in a homogeneous well-mixed population \cite{Wang2024}, but it remains challenging to comprehensively analyze how risk preferences affect cooperation in heterogeneous social networks. 

Our work provides a systematic framework to address this problem, by taking arbitrary distributions of risk preferences and arbitrary network structures into account.
We find that, at the population level, risk-averse behaviors foster the evolution of cooperation and even enable the emergence of cooperation in some graphs where cooperation would fail to evolve in the traditional model. At the node level, nodes with high degrees are more effective at learning and propagating their strategies. Furthermore, when node degrees are identical, nodes with lower betweenness centrality dominate the direction of evolution. 
As a result, if such nodes are risk-averse, cooperation spreads more readily within local clusters, thereby promoting its emergence across the entire network.

In our model, for analytical tractability, the function of the expected utility with respect to risk preferences and punishment probability (Fig.~\ref{fig:1}c) is modeled as a piecewise linear function. A natural and realistic extension is to model it as a smooth function: it is convex when the risk preference is positive and is concave when it is negative (see Supplementary Fig.~1{a}), which preserves the key property of the piecewise linear formulation. 
We conducted numerical simulations on complete graphs, scale-free networks, and regular networks (see Supplementary Fig.~1{b}-1{d}) to validate the effect of mean risk preference, the joint effect of degree variance and risk preference variance, and the role of betweenness centrality on strategy evolution, respectively, under the extended model. 
The simulations are consistent with the theoretical predictions obtained under the piecewise linear model.

We have proposed a mathematical framework to investigate the evolution of cooperation when the payoff is uncertain and all individuals have risk preferences. Nevertheless, several limitations remain to be addressed. First, our model focuses exclusively on pairwise interactions, whereas real-world interactions often involve more than two participants simultaneously \cite{Santos2008,Perc2013,Gokhale2014,Mullon2014,Pena2015,VanCleve2015,Benson2016,Battiston2020,Battiston2021,sheng2024strategy}. Such higher-order interactions give rise to multi-player game structures, such as public goods games \cite{Hauert2006,Santos2008}, which cannot be adequately captured by the donation game. How to incorporate risk preferences and uncertainty into multi-player settings remains an open avenue for future research. 
Second, we model uncertainty through probabilistic punishment, but noise in real-world social interactions is often multifaceted. Factors such as random fluctuations in individual decision-making and the exploration of novel strategies \cite{traulsen2009exploration} cannot be simply modeled by Bernoulli random variables in our model. How to incorporate uncertainty arising from other social behaviors, such as white noise and colored noise, into social networks is still an open but practical question. These challenges represent promising directions for future research on strategic interactions under risk preferences and uncertainty.

\section*{Methods}
\paragraph{Fixation probability under risk preference.}
We use the workflow proposed by McAvoy et al. \cite{mcavoy2021fixation} to derive the analytical form of the fixation probability for cooperation. We refer to the Supplementary Information for more details.

Let $\pi_i=\sum_{j=1}^N w_{ij}/\sum_{i,j=1}^N w_{ij}$ denote the reproductive value of individual $i$, and let \(p_{ij}^{(k)}\) denote the probability of a \(k\)-step random-walk from \(i\) to \(j\).
We rewrite the payoff of individual $j$ from $i$'s perspective, denoted $u_{ij}$ (Eq.~\ref{eq:uij}), as
\begin{equation}
    u_{ij}=\sum_{I\subseteq\{1,\dots,N\}} d_I^{ij}\,\mathbf{x}_I,
\end{equation}
where \(\mathbf{x}_I=\prod_{k\in I} x_k\) and \(x_k\in\{0,1\}\) denote the state of individual $k$, and $|I|$ takes values in $\{1,2,3\}$.
Then the fixation probability for cooperation under weak selection (\(\delta\to 0\)) is
\begin{equation}
    \rho_{C}=\frac{1}{N}+\frac{\delta }{N} \left[\sum_{i,j,l=1}^{N}\sum_{\substack{I \subseteq \{1,...,N\} \\ 0 \le |I| \le 3}} \pi_{i}p_{ij}^{(1)}\left ( p_{il}^{(1)} -p_{jl}^{(0)}\right )\eta _{\left \{ j\right \} \cup I }d_{I}^{il}  \right ]+O(\delta^{2}),
    \label{eq:fix_p}
\end{equation}
The quantities \(\eta_{I}\) are uniquely determined by the following linear system
\begin{equation}
    \begin{aligned}
       \eta_{i_{1}\cdots i_{l}} &= \frac{1}{l}+\frac{1}{l} \sum_{y=1}^{N}\left(   p_{i_{1}y}^{(1)} \eta_{y\cdots i_{l}} + \dots +   p_{i_{l}y}^{(1)} \eta_{i_{1}\cdots y}\right) \quad 2 \le l \le 4; \\
       \eta_{\left\{ i \right\} } &= 0 \quad l = 1.
    \end{aligned}
\end{equation}

Substituting Eq.~\ref{eq:uij} into Eq.~\ref{eq:fix_p} gives
\begin{equation}
    \rho_{C} = \frac{1}{N} + \frac{\delta}{N} \left[ \left(f_{b}^{(1)}+f_{b}^{(2)} \right) b - \left(f_{c}^{(1)}+f_{c}^{(2)} \right) c \right] + O(\delta^{2}),
\end{equation}
where
\begin{equation}
    \begin{aligned}
        f_{c}^{(1)} &= \sum_{i,j=1}^{N} \pi_i p_{ij}^{(2)} \eta_{ij},\\ 
    f_{c}^{(2)} &= Qp\sum_{i,j=1}^{N} \pi_i p_{ij}^{(2)} \eta_{ij},\\
    f_{b}^{(1)} &= \sum_{i,j=1}^{N} \pi_i \left(p_{ij}^{(3)}-p_{ij}^{(1)} \right) \eta_{ij}, \\ 
    f_{b}^{(2)} &= \sum_{i,j,l,j_{1}=1}^{N} \pi_{i} p_{ij}^{(1)} p_{il}^{(1)} p_{lj_{1}}^{(1)} p_{lj_{1}}^{(1)} \alpha_{i} B \left( \eta_{jlj_{1}} - \eta_{jj_{1}} \right) \\
    &\quad + \sum_{i,j,l,j_{1},j_{2}=1}^{N} \pi_{i} p_{ij}^{(1)} p_{il}^{(1)} p_{lj_{1}}^{(1)} p_{lj_{2}}^{(1)} \alpha_{i} B \left( \eta_{jlj_{1}j_{2}} - \eta_{jj_{1}j_{2}} \right),
    \end{aligned}
\end{equation}
and 
\begin{equation}
    \alpha_{i}=\begin{cases}
        p(1-A_{i}) & p\le0.5,\\
        1-(1-p)(A_{i}+1) & p\ge0.5. \\
    \end{cases}
\end{equation}

\paragraph{Definition of betweenness centrality.}
The betweenness centrality quantifies how often a node lies on shortest paths between other nodes in a graph. In a weighted graph, a shortest path between nodes $s$ and $t$ is defined as a path that minimizes the sum of the weights of its constituent edges. Formally, for a graph $G = (V, E)$, let $\sigma_{st}$ denote the total number of shortest paths between two distinct nodes $s, t \in V$, and let $\sigma_{st}(v)$ denote the number of those paths that pass through a node $v \neq s, t$. The betweenness centrality of node $v$ is then defined as
\begin{equation}
    C_B(v) = \sum_{s \neq t \neq v} \frac{\sigma_{st}(v)}{\sigma_{st}}.   
\end{equation}

\bibliographystyle{unsrt}
\bibliography{ref}

\newpage


\section*{Supplementary material}

# 1  Model description

We consider the evolutionary dynamics on a population of $N$ individuals, labeled $\mathcal{N} = \{1, ..., N\}$. The population structure is represented by a graph, in which nodes correspond to individuals and edges indicate pairwise interactions. We use an adjacency matrix $W = (w_{ij})_{i,j=1}^{N}$ to capture the graph structure, where the entry $w_{ij}$ denotes the interaction weight from individual $i$ towards individual $j$. We further assume that the graph is undirected ($w_{ij} = w_{ji}$ for all $i \neq j$) and contains no self-loops ($w_{ii} = 0$ for all $i$).

Let $p_{ij}^{(m)}$ denote the probability that a random walk started at node $i$ reaches node $j$ in $m$ steps, and let $\pi_i$ be the reproductive value of node $i$ [1]. We have $p_{ij}^{(1)} = w_{ij}/\sum_{j=1}^{N} w_{ij}$, $p_{ij}^{(m)} = \left(p^{(1)}\right)_{ij}^{m}$ and $\pi_i = \sum_{j=1}^{N} w_{ij} / \sum_{i,j=1}^{N} w_{ij}$.

At each time step, each individual adopts one of two strategies: cooperation (C) or defection (D) and engages in pairwise interactions with their neighbors. Let $x_i$ denote the strategy of the individual $i$, where $x_i = 1$ if the individual $i$ is cooperator and $x_i = 0$ otherwise, and let $\mathbf{x} = (x_1, ..., x_N)$ denote the state of the population.

Each individual $i$ initially receives an average payoff $v_i^{(0)}$ over all possible interactions. Then, defector $i$ might be punished with probability $p$, which reduces their payoff to a fraction $\beta_i$ of the original value, and cooperators pay an additional cost $\alpha$ to punish defectors. Let $u_i^{(0)}$ and $u_i^{(1)}$ denote the rewards of individual $i$ before and after punishment, respectively. We have

$$
\begin{aligned}
u_i^{(0)} &= v_i^{(0)} - \alpha x_i, \\
u_i^{(1)} &= x_i u_i^{(0)} + (1 - x_i)\beta_i u_i^{(0)}.
\end{aligned}
\qquad (1)
$$

The final payoff of $i$ is then a random variable

$$
u_i = u_i^{(1)} X + u_i^{(0)}(1 - X), \qquad (2)
$$

where $X$ follows a Bernoulli distribution with parameter $p$.

Individuals perceive this randomness in different ways based on their own risk preferences. Let $A_i$ denote the risk preference of $i$. The payoff of individual $j$ from $i$'s perspective is

$$
u_{ij} = \begin{cases} u_j^{(0)} - p(1 - A_i)\left(u_j^{(0)} - u_j^{(1)}\right) & p \leq 0.5, \\ u_j^{(1)} + (1 - p)(A_i + 1)\left(u_j^{(0)} - u_j^{(1)}\right) & p \geq 0.5. \end{cases} \qquad (3)
$$

When $A_i = 0$, individual $i$ is risk-neutral, and the payoff $u_{ij}$ is the expectation of $u_j$ (i.e. the $u_{ij}$-$p$ curve is linear). When $A_i > 0$ (resp. $A_i < 0$), individual $i$ is risk-seeking (resp. risk-averse), making the curve $u_{ij}$-$p$ convex (resp. concave).

At the end of each time step, a randomly chosen individual $i$ updates their strategy by imitating one of their neighbors. The probability that $i$ copies $j$'s strategy is proportional

to the edge weight $w_{ij}$ and $j$'s fitness from $i$'s perspective, $f_{ij} = 1 + \delta u_{ij}$, where $\delta$ indicates the intensity of selection.

## 2 Results

In this study, we consider the donation game, where a cooperator pays a cost of $c$ to provide a benefit of $b$ to their opponents, regardless of their opponents' strategies, and a defector pays no cost and provides no benefit. Then, the initial payoff of individual $i$ is

$$v_i^{(0)} = -cx_i + b\sum_{j=1}^{N} p_{ij}^{(1)} x_j. \tag{4}$$

The additional cost $\alpha$ scales with the initial cost $c$ and the probability of punishment $p$, and is defined as $\alpha = cpQ$, where $Q \in [0,1]$ is a free parameter. The strength of punishment $\beta_i$ is positively related to the number of cooperative neighbors, defined as $\beta_i = 1 - \sum_{j=1}^{N} p_{ij}^{(1)} x_j B$, where $B \in [0,1]$ is a free parameter. In this setting, the payoffs of individual $i$ before and after punishment become

$$\begin{aligned} u_i^{(0)} &= -cx_i + b\sum_{j=1}^{N} p_{ij}^{(1)} x_j - cpQx_i, \\ u_i^{(1)} &= \left[ -cx_i + b\sum_{j=1}^{N} p_{ij}^{(1)} x_j - cQpx_i \right] \left[ 1 - (1-x_i)\sum_{j=1}^{N} p_{ij}^{(1)} x_j B \right]. \end{aligned} \tag{5}$$

We rewrite the payoffs in Eq. 5 as multi-linear polynomial functions,

$$u_i^{(0)} = \sum_{I \subseteq \mathcal{N}} c_I^{i,0} x_I, \; u_i^{(1)} = \sum_{I \subseteq \mathcal{N}} c_I^{i,1} x_I, \tag{6}$$

where $c_I^{i,0}$ and $c_I^{i,1}$ are coefficients indexed by the subsets $I \subseteq \mathcal{N}$,

$$\begin{aligned} c_i^{i,0} &= -c(1+Qp), \\ c_k^{i,0} &= bp_{ik}^{(1)}. \\ c_i^{i,1} &= -c(1+Qp), \\ c_k^{i,1} &= bp_{ik}^{(1)}(1 - p_{ik}^{(1)} B), \\ c_{ik}^{i,1} &= bp_{ik}^{(1)} p_{ik}^{(1)} B, \\ c_{kl}^{i,1} &= -2bp_{ik}^{(1)} p_{il}^{(1)} B, \\ c_{ikl}^{i,1} &= 2bp_{ik}^{(1)} p_{il}^{(1)} B, \end{aligned} \tag{7}$$

and $x_I = \prod_{i \in I} x_i$.

Thus, the payoff of individual $j$ from $i$'s perspective can be written as

$$u_{ij} = \sum_{I \subseteq \mathcal{N}} d_I^{ij} x_I \tag{8}$$

where the coefficients are

$$d_I^{ij} = \begin{cases} [1 - p(1 - A_i)] c_I^{j,0} + p(1 - A_i) c_I^{j,1} & p \leq 0.5, \\ (1-p)(A_i + 1) c_I^{j,0} + [1 - (1-p)(A_i + 1)]) c_I^{j,1} & p \geq 0.5. \end{cases} \tag{9}$$

Plugging Eq. 7 into Eq. 9 gives

$$\begin{aligned} d_j^{ij} &= -c\left(1 + Qp\right), \\ d_k^{ij} &= b p_{jk}^{(1)} (1 - \alpha_i p_{jk}^{(1)} B), \\ d_{jk}^{ij} &= \alpha_i b p_{jk}^{(1)} p_{jk}^{(1)} B, \\ d_{kl}^{ij} &= -2\alpha_i b p_{jk}^{(1)} p_{jl}^{(1)} B, \\ d_{jkl}^{ij} &= 2\alpha_i b p_{jk}^{(1)} p_{jl}^{(1)} B. \end{aligned} \tag{10}$$

where the parameter $\alpha_i$ is defined as

$$\alpha_i = \begin{cases} p(1 - A_i) & p \leq 0.5, \\ 1 - (1-p)(A_i + 1) & p \geq 0.5. \end{cases} \tag{11}$$

We can observe that $\alpha_i \in [0, 1]$.

## 2.1 Fixation probability

We now proceed to calculate the fixation probability of cooperation within the framework proposed by McAvoy and Allen [2]. First, we derive several necessary quantities. Under the death-birth process, the probability of a replacement event $(\{j\}, j \to i)$ occurring in state $\mathbf{x}$ is given by

$$p_{(\{j\}, j \to i)}(\mathbf{x}) = \frac{1}{N} \frac{w_{ji} f_{ji}(\mathbf{x})}{\sum_{l=1}^{N} w_{jl} f_{jl}(\mathbf{x})}, \tag{12}$$

Under neutral drift ($\delta = 0$), we have $p_{(\{j\}, j \to i)}^{\circ}(\mathbf{x}) = \frac{p_{ji}^{(1)}}{N}$. Therefore, the marginal probability that individual $i$ transmits its strategy to individual $j$ under weak selection ($\delta \ll 1$) is given by

$$e_{ij}(\mathbf{x}) = e_{ij}(f_j(\mathbf{x})) = \frac{1}{N} \frac{f_{ji} w_{ji}}{\sum_{l=1}^{N} f_{jl} w_{jl}} = \frac{1}{N} \frac{(1 + \delta u_{ji}) w_{ji}}{\sum_{l=1}^{N} (1 + \delta u_{jl}) w_{jl}}, \tag{13}$$

where $f_j(\mathbf{x}) = (f_{j1}, \ldots, f_{jN})$ denotes the vector of fitness values that individual $j$ assigns to other individuals. Differentiating the above expression with respect to $\delta$ yields

$$\frac{d}{d\delta}\bigg|_{\delta=0} e_{ij}(\mathbf{x}) = \frac{p_{ji}^{(1)}}{N}\left[\sum_{l=1}^{N}\left(p_{il}^{(0)} - p_{jl}^{(1)}\right)u_{jl}\right] = \sum_{I \subseteq \mathcal{N}} b_I^{ij} x_I. \tag{14}$$

Therefore, the coefficient $b_I^{ij}$ can be expressed as

$$b_I^{ij} = \frac{p_{ji}^{(1)}}{N}\left[\sum_{l=1}^{N}\left(p_{il}^{(0)} - p_{jl}^{(1)}\right)d_I^{jl}\right], \tag{15}$$

where $1 \leq |I| \leq 3$.

Under weak selection, when the initial configuration is uniform, that is, when a single cooperator is randomly introduced into an otherwise all-defector population, the fixation probability of cooperation becomes

$$\rho_C = \frac{1}{N} + \frac{\delta}{N}\left[\sum_{i,j,l=1}^{N}\sum_{\substack{I \subseteq \mathcal{N} \\ 0 \leq |I| \leq 3}} \pi_i p_{ij}^{(1)}\left(p_{il}^{(1)} - p_{jl}^{(0)}\right)\eta_{\{j\}\cup I}d_I^{il}\right] + O(\delta^2), \tag{16}$$

where the coefficients $\eta_I$ are given as the unique solution to the following system of linear equations,

$$\eta_{i_1 \cdots i_l} = \frac{1}{l} + \frac{1}{l}\sum_{y=1}^{N}\left(p_{i_1 y}^{(1)}\eta_{y \cdots i_l} + \cdots + p_{i_l y}^{(1)}\eta_{i_1 \cdots y}\right) \quad 2 \leq l \leq 4;$$
$$\eta_{\{i\}} = 0 \quad l = 1. \tag{17}$$

The first-order term in Eq. (16) is

$$\sum_{i,j,l=1}^{N}\sum_{\substack{I \subseteq \mathcal{N} \\ 0 \leq |I| \leq 3}} \pi_i p_{ij}^{(1)}\left(p_{il}^{(1)} - p_{jl}^{(0)}\right)\eta_{\{j\}\cup I}d_I^{il}, \tag{18}$$

which separates into three cases based on the value of $|I|$. For simplicity, we define

$$S_{ijlI} := \pi_i p_{ij}^{(1)}\left(p_{il}^{(1)} - p_{jl}^{(0)}\right)\eta_{\{j\}\cup I}d_I^{il}. \tag{19}$$

1. When $|I| = 1$, we have

$$\sum_{j_1=1}^{N} \sum_{i,j,l=1}^{N} = \sum_{i,j,l=1}^{N} \left( \sum_{j_1=l}^{N} S_{ijlj_1} + \sum_{j_1 \neq l}^{N} S_{ijlj_1} \right)$$
$$= -c \sum_{i,j,l=1}^{N} (1 + Qp) \pi_i p_{ij}^{(1)} \left( p_{il}^{(1)} - p_{jl}^{(0)} \right) \eta_{jl} \qquad , \qquad (20)$$
$$+ \sum_{i,j,l,j_1=1}^{N} \pi_i p_{ij}^{(1)} \left( p_{il}^{(1)} - p_{jl}^{(0)} \right) \eta_{jj_1} b p_{lj_1}^{(1)} \left( 1 - \alpha_i p_{lj_1}^{(1)} B \right)$$

2. When $|I| = 2$, we have

$$\sum_{j_1,j_2=1}^{N} \sum_{i,j,l=1}^{N} S_{ijlj_1j_2} = \sum_{i,j,l=1}^{N} \left( \sum_{\substack{j_1,j_2=1 \\ j_1 \neq j_2, l \in \{j_1,j_2\}}}^{N} S_{ijlj_1j_2} + \sum_{\substack{j_1,j_2=1 \\ j_1 \neq j_2, l \notin \{j_1,j_2\}}}^{N} S_{ijlj_1j_2} \right)$$
$$= \sum_{i,j,l,j_1=1}^{N} \pi_i p_{ij}^{(1)} \left( p_{il}^{(1)} - p_{jl}^{(0)} \right) \eta_{jlj_1} \alpha_i b p_{lj_1}^{(1)} p_{lj_1}^{(1)} B \qquad , \qquad (21)$$
$$- \sum_{i,j,l,j_1,j_2=1}^{N} \pi_i p_{ij}^{(1)} \left( p_{il}^{(1)} - p_{jl}^{(0)} \right) \eta_{jj_1j_2} \alpha_i b p_{lj_1}^{(1)} p_{lj_2}^{(1)} B$$

3. When $|I| = 3$, we have

$$\sum_{\substack{I \subseteq \mathcal{N} \\ |I|=3}} \sum_{i,j,l=1}^{N} S_{ijlI} = \sum_{i,j,l=1}^{N} \left( \sum_{\substack{I \subseteq \mathcal{N} \\ |I|=3, l \in I}} S_{ijlI} + \sum_{\substack{I \subseteq \mathcal{N} \\ |I|=3, l \notin I}} S_{ijlI} \right) \qquad . \qquad (22)$$
$$= \sum_{i,j,l,j_1,j_2=1}^{N} \pi_i p_{ij}^{(1)} \left( p_{il}^{(1)} - p_{jl}^{(0)} \right) \eta_{jlj_1j_2} \alpha_i b p_{lj_1}^{(1)} p_{lj_2}^{(1)} B$$

Then the first-order term becomes

$$\sum_{i,j,l=1}^{N} \sum_{\substack{I \subseteq \mathcal{N} \\ 0 \leq |I| \leq 3}} \pi_i p_{ij}^{(1)} \left( p_{il}^{(1)} - p_{jl}^{(0)} \right) \eta_{\{j\} \cup I} c_I^{il} := f_b b - f_c c, \qquad (23)$$

where

$$f_c = \sum_{i,j=1}^{N} (1 + Qp) \pi_i p_{ij}^{(2)} \eta_{ij}, \qquad (24)$$

6

$$f_b = \sum_{i,j,l,j_1=1}^{N} \pi_i p_{ij}^{(1)} \left(p_{il}^{(1)} - p_{jl}^{(0)}\right) \eta_{jj_1} p_{lj_1}^{(1)} \left(1 - \alpha_i p_{lj_1}^{(1)} B\right)$$
$$+ \sum_{i,j,l,j_1=1}^{N} \pi_i p_{ij}^{(1)} \left(p_{il}^{(1)} - p_{jl}^{(0)}\right) \eta_{jlj_1} \alpha_i p_{lj_1}^{(1)} p_{lj_1}^{(1)} B$$
$$- \sum_{i,j,l,j_1,j_2=1}^{N} \pi_i p_{ij}^{(1)} \left(p_{il}^{(1)} - p_{jl}^{(0)}\right) \eta_{jj_1j_2} \alpha_i p_{lj_1}^{(1)} p_{lj_2}^{(1)} B$$
$$+ \sum_{i,j,l,j_1,j_2=1}^{N} \pi_i p_{ij}^{(1)} \left(p_{il}^{(1)} - p_{jl}^{(0)}\right) \eta_{jlj_1j_2} \alpha_i p_{lj_1}^{(1)} p_{lj_2}^{(1)} B. \tag{25}$$

We define $\eta_{(n)} = \sum_{i,j=1}^{N} \pi_i p_{ij}^{(n)} \eta_{ij}$. When the network contains no self-loops, we can further simplify $f_c$ and $f_b$,

$$f_c = (1 + Qp)\eta_{(2)} = f_c^{(1)} + f_c^{(2)}, \tag{26}$$

and

$$f_b = \eta_{(3)} - \eta_{(1)}$$
$$+ \sum_{i,j,l,j_1=1}^{N} \pi_i p_{ij}^{(1)} p_{il}^{(1)} p_{lj_1}^{(1)} p_{lj_1}^{(1)} \alpha_i B \left(\eta_{jlj_1} - \eta_{jj_1}\right)$$
$$+ \sum_{i,j,l,j_1,j_2=1}^{N} \pi_i p_{ij}^{(1)} p_{il}^{(1)} p_{lj_1}^{(1)} p_{lj_2}^{(1)} \alpha_i B (\eta_{jlj_1j_2} - \eta_{jj_1j_2}) \tag{27}$$
$$= f_b^{(1)} + f_b^{(2)}.$$

where $f_b^{(1)} = \eta_{(3)} - \eta_{(1)}$ ($f_c^{(1)} = \eta_{(2)}$) denotes the benefit (cost) derived under the classical framework, and $f_b^{(2)}$ ($f_c^{(2)}$) captures the additional benefit (cost) arising from risk preferences and payoff uncertainty.

The critical benefit-to-cost ratio [3], denoted by $(b/c)^*$, is given by

$$\left(\frac{b}{c}\right)^* = \frac{f_c^{(1)} + f_c^{(2)}}{f_b^{(1)} + f_b^{(2)}}. \tag{28}$$

## 2.2 Penalty-cost function

In this section, our goal is to clarify the relationship between the cost paid by cooperators and the punishment imposed on defectors, under the constraint that the critical ratio remains unchanged. Establishing such a relationship enables us to further investigate how the punishment cost changes with the intensity of punishment on a given network.

In the present model, the parameter $Q$ measures the cost paid by cooperators, while the parameters $B$ and $p$ characterize the magnitude and probability of punishment experienced by defectors. Accordingly, we fix the critical ratio $(b/c)^*$ and, based on this constraint, derive the functional relationship between $Q$ and the pair $(B, p)$.

When $(b/c)^*$ is fixed at a value $r^*$, we define

$$X_{ijlj_1} = \pi_i p_{ij}^{(1)} p_{il}^{(1)} p_{lj_1}^{(1)} \left[ \eta_{jlj_1} p_{lj_1}^{(1)} - \eta_{jj_1} p_{lj_1}^{(1)} + \sum_{j_2=1}^{N} (\eta_{jlj_1j_2} - \eta_{jj_1j_2}) p_{lj_2}^{(1)} \right], \quad (29)$$

where $\Psi_{ijlj_1}$ only depends on the network structure and nodes $i, j, l, j_1$. By substituting the above parameters into Eq. (28) and solving, we obtain

$$Q = \begin{cases} \dfrac{1}{\eta_{(2)}p} \left[ r^* \left( \eta_{(3)} - \eta_{(1)} \right) + r^* B p \sum_{i,j,l,j_1=1}^{N} (1 - A_i) X_{ijlj_1} - \eta_{(2)} \right] & p \leq 0.5, \\ \dfrac{1}{\eta_{(2)}p} \left[ r^* \left( \eta_{(3)} - \eta_{(1)} \right) + r^* B \sum_{i,j,l,j_1=1}^{N} [1 - (1-p)(1 + A_i)] X_{ijlj_1} - \eta_{(2)} \right] & p \geq 0.5. \end{cases} \quad (30)$$

Therefore, theoretically, this equation establishes a connection between the intensity of punishment and its associated cost.

## 3 Applications to specific networks

### 3.1 Well-mixed populations

Firstly, we consider a well-mixed population of size $N$. We have

$$\begin{aligned} p_{ij}^{(1)} &= \frac{1}{N-1}, p_{ii}^{(1)} = 0, \\ p_{ij}^{(2)} &= \frac{N-2}{(N-1)^2}, p_{ii}^{(2)} = \frac{1}{N-1}, \\ p_{ij}^{(3)} &= \frac{N^2 - 3N + 3}{(N-1)^3}. \end{aligned} \quad (31)$$

Due to symmetry, each node in a complete graph possess identical properties. Acoordingly, we have

$$\begin{cases} \eta_i = 0, \\ \eta_{i_1 i_2} = \frac{N-1}{2}, \\ \eta_{i_1 i_2 i_3} = \frac{2(N-1)}{3}, \\ \eta_{i_1 i_2 i_3 i_4} = \frac{3(N-1)}{4}, \end{cases} \quad (32)$$

and

$$\begin{cases} \eta_{(1)} = \sum_{\substack{i,j=1 \\ i \neq j}}^{N} \pi_i p_{ij}^{(1)} \eta_{ij} = \dfrac{N-1}{2}, \\ \eta_{(2)} = \sum_{\substack{i,j=1 \\ i \neq j}}^{N} \pi_i p_{ij}^{(2)} \eta_{ij} = \dfrac{N-2}{2}, \\ \eta_{(3)} = \sum_{\substack{i,j=1 \\ i \neq j}}^{N} \pi_i p_{ij}^{(3)} \eta_{ij} = \dfrac{N^2 - 3N + 3}{2(N-1)}. \end{cases} \qquad (33)$$

Plugging Eqs. 32 and 33 into Eqs. 26 and 27 gives

$$f_c = (1 + Qp)\frac{N-2}{2}, \qquad (34)$$

and

$$f_b = \frac{2-N}{2(N-1)} + \frac{B(N-2)(N+1)}{12(N-1)^2} \sum_{i=1}^{N} \alpha_i. \qquad (35)$$

Therefore, the fixation probability in the well-mixed population is given by

$$\rho_A = \frac{1}{N} + \frac{\delta(N-2)}{2N} \left\{ b \left[ \frac{B(N+1)}{6(N-1)^2} \sum_{i=1}^{N} \alpha_i - \frac{1}{N-1} \right] - c(1 + Qp) \right\}, \qquad (36)$$

and the critical benefit-to-cost ratio is

$$\left(\frac{b}{c}\right)^* = \frac{6(1 + Qp)(N-1)^2}{B(N+1)\sum_{i=1}^{N} \alpha_i - 6(N-1)}. \qquad (37)$$

We now proceed to calculate the admissible range of $(b/c)^*$. It is evident that $6(1 + Qp)(N-1)^2 > 0$ always holds; therefore, the sign of Eq. (37) depends solely on the denominator.

When $N$ is sufficiently small such that $BN(N+1) < 6(N-1)$, the inequality $(b/c)^* < 0$ always holds. The miniimum value of $12(N-1)^2/[BN(N+1) - 6(N-1)]$ is attained when $p = 1$ and $Q = 1$, whereas the maximum value $-(N-1)$ occurs when $p = 0$. Therefore, the admissible range of $(b/c)^*$ is given by

$$\left(\frac{b}{c}\right)^* \in \left[\frac{12(N-1)^2}{BN(N+1) - 6(N-1)}, -(N-1)\right]. \qquad (38)$$

When $N$ is sufficiently large such that $BN(N+1) > 6(N-1)$. When $(b/c)^* < 0$, the maximum value $-(N-1)$ is attained when $p = 0$. When $(b/c)^* > 0$, the minimum value $6(N-1)^2/[BN(N+1) - 6(N-1)]$ is attained when $Q = 0$ and $p = 0.5$ and $A_i = -1$

$(1 \leq i \leq N)$. Therefore, the admissible range of $(b/c)^*$ is given by

$$\left(\frac{b}{c}\right)^* \in (-\infty, -(N-1)] \cup \left[\frac{6(N-1)^2}{BN(N+1) - 6(N-1)}, +\infty\right). \quad (39)$$

When $N \to \infty$, the admissible range of $(b/c)^*$ is given by

$$\left(\frac{b}{c}\right)^* \in \{-\infty\} \cup \left[\frac{6}{B}, +\infty\right). \quad (40)$$

## 3.2 Star networks

Secondly, we consider star networks. There are two types of nodes in the star network. The central node, called the root node, is denoted by $r$; the remaining nodes are leaf nodes, denoted by $l$. Let $n$ denote the number of leaf nodes.

When $|I| = 2$, only two variables exist, $\eta_{rl}$ and $\eta_{l_1 l_2}$, which represent the pairs consisting of the root and a leaf node, and two distinct leaf nodes, respectively. Thus, the linear system reduces to

$$\eta_{rl} = \frac{1}{2} + \frac{1}{2}\left(\frac{n-1}{n}\eta_{l_1 l_2}\right);$$
$$\eta_{l_1 l_2} = \frac{1}{2} + \eta_{rl}.$$

Solving the system gives

$$\eta_{rl} = \frac{3n-1}{2(n+1)}, \eta_{l_1 l_2} = \frac{2n}{n+1}. \quad (41)$$

When $|I| = 3$, the linear system has two variables, $\eta_{rl_1 l_2}$ and $\eta_{l_1 l_2 l_3}$, which represent the sets consisting of the root node and two distinct leaf nodes, and three distinct leaf nodes, respectively. Therefore, the linear system becomes

$$\eta_{rl_1 l_2} = \frac{1}{3} + \frac{1}{3}\left(\frac{n-2}{n}\eta_{l_1 l_2 l_3} + \frac{2}{n}\eta_{l_1 l_2} + 2\eta_{rl}\right);$$
$$\eta_{l_1 l_2 l_3} = \frac{1}{3} + \eta_{rl_1 l_2}.$$

Solving the system gives

$$\eta_{rl_1 l_2} = \frac{13n-2}{6(n+1)}, \eta_{l_1 l_2 l_3} = \frac{5n}{2(n+1)}. \quad (42)$$

When $|I| = 4$, the linear system contains two variables, $\eta_{rl_1 l_2 l_3}$ and $\eta_{l_1 l_2 l_3 l_4}$, which correspond to the root node with three distinct leaf nodes, and four distinct leaf nodes,

respectively. Hence, the linear system becomes

$$\eta_{rl_1l_2l_3} = \frac{1}{4} + \frac{1}{4}\left(\frac{n-3}{n}\eta_{l_1l_2l_3l_4} + \frac{3}{n}\eta_{l_1l_2l_3} + 3\eta_{rl_1l_2}\right);$$
$$\eta_{l_1l_2l_3l_4} = \frac{1}{4} + \eta_{rl_1l_2l_3}.$$

Solving the system gives

$$\eta_{rl_1l_2l_3} = \frac{31n-3}{12(n+1)}, \eta_{l_1l_2l_3l_4} = \frac{17n}{6(n+1)}. \tag{43}$$

The first-order transition probabilities are given by

$$\begin{aligned}
p_{rl}^{(1)} &= \frac{1}{n}, \quad p_{rr}^{(1)} = 0, \\
p_{lr}^{(1)} &= 1, \quad p_{l_1l_1}^{(1)} = 0, \quad p_{l_1l_2} = 0,
\end{aligned} \tag{44}$$

where $l_1l_1$ denotes two identical leaf nodes, and $l_1l_2$ denotes two distinct leaf nodes.

The second-order transition probabilities are given by

$$\begin{aligned}
p_{rl}^{(2)} &= 0, \quad p_{rr}^{(2)} = 1, \\
p_{lr}^{(2)} &= 0, \quad p_{l_1l_1}^{(2)} = \frac{1}{n}, \quad p_{l_1l_2}^{(2)} = \frac{1}{n},
\end{aligned} \tag{45}$$

The third-order transition probabilities are given by

$$\begin{aligned}
p_{rl}^{(3)} &= \frac{1}{n}, \quad p_{rr}^{(3)} = 0, \\
p_{lr}^{(3)} &= 1, \quad p_{l_1l_1}^{(3)} = 0, \quad p_{l_1l_2}^{(3)} = 0,
\end{aligned} \tag{46}$$

It is worth noting that for any two nodes, the first-order transition probability $p_{ij}^{(1)}$ equals the third-order transition probability $p_{ij}^{(3)}$. Similarly, by mathematical induction, this result can be generalized to the relationship between the $k$-th and $k+2$-th order transition probabilities:

$$p_{ij}^{(k)} = p_{ij}^{(k+2)}. \tag{47}$$

In particular, we have

$$\begin{cases} \eta_{(1)} = \sum_{\substack{i,j=1 \\ i\neq j}}^{N} \eta_{ij} = \dfrac{n+1}{2n}\eta_{rl} = \dfrac{3n-1}{4n}, \\ \eta_{(2)} = \sum_{\substack{i,j=1 \\ i\neq j}}^{N} \pi_i p_{ij}^{(2)} \eta_{ij} = \dfrac{n-1}{2n}\eta_{l_1 l_2} = \dfrac{n-1}{n+1}, \\ \eta_{(3)} = \sum_{\substack{i,j=1 \\ i\neq j}}^{N} \pi_i p_{ij}^{(3)} \eta_{ij} = \sum_{i,j;i\neq j}^{N} \pi_i p_{ij}^{(3)} \eta_{ij} = \dfrac{3n-1}{4n}, \end{cases} \quad (48)$$

and

$$f_c = (1+Qp)\frac{n-1}{n+1}, \quad (49)$$

$$f_b = B\alpha_r \frac{(n-1)(4n+1)}{12n(n+1)}. \quad (50)$$

Therefore, the fixation probability in the star network is given by

$$\rho_C = \frac{1}{n+1} + \frac{(n-1)\delta}{(n+1)^2}\left\{bB\alpha_r \frac{4n+1}{12n} - c(1+Qp)\right\}, \quad (51)$$

and the critical benefit-to-cost ratio is

$$\left(\frac{b}{c}\right)^* = \frac{12n(1+Qp)}{(4n+1)B\alpha_r}. \quad (52)$$

It is clear that $\alpha_r \geq 0$ always holds. Therefore, $(b/c)^*$ is always positive. The maximum value, $+\infty$, is attained when $p = 0$, whereas the minimum value, $12n/(4n+1)$, occurs when $Q = 0$ and $B = 1$ and $p = 0.5$ and $A_r = -1$. Therefore, the admissible range of $(b/c)^*$ is given by

$$(b/c)^* \in \left[\frac{12n}{4n+1}, +\infty\right). \quad (53)$$

When $n \to \infty$, the admissible range of $(b/c)^*$ is given by

$$(b/c)^* \in [3, +\infty). \quad (54)$$

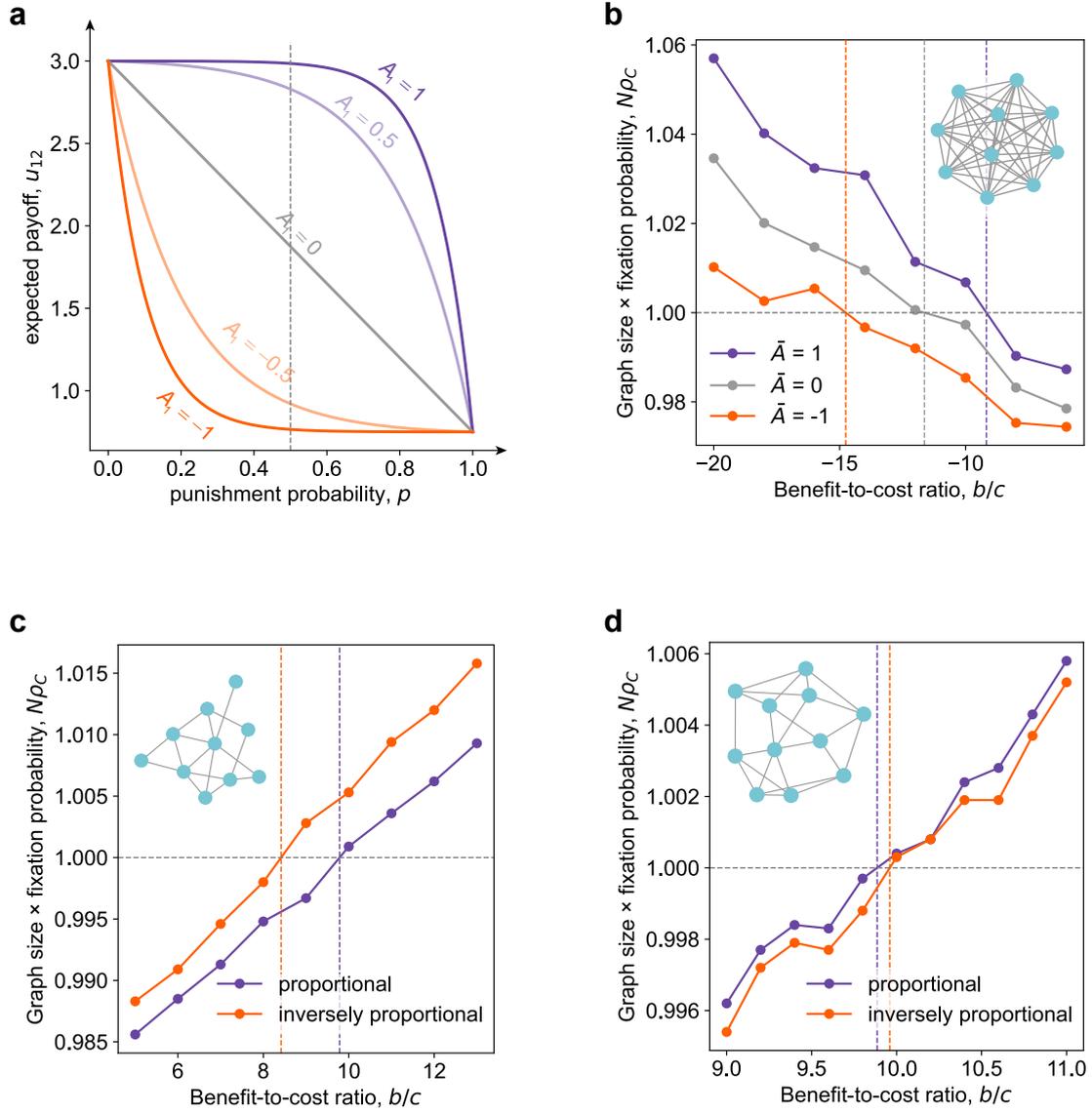

**Supplementary Figure 1: Results under continuous utility functions.** (**a**) Illustration of the piecewise-defined utility function $u_{ij}$ for different values of $A_i$ (1, 0.5, 0, −0.5, and −1), where $A_i = 0$ corresponds to $u_{ij} = u_j^{(0)} + (u_j^{(1)} - u_j^{(0)})p$, and $A_i \neq 0$ corresponds to $u_{ij} = u_j^{(0)} + (u_j^{(1)} - u_j^{(0)}) \frac{1-e^{10Ap}}{1-e^{10A}}$. The curvature of the function remains consistent with that of the original utility function for different $A_i$ values. (**b**) Simulation results for a complete graph with $N = 10$ nodes, where each point represents $10^6$ Monte Carlo realizations. The critical benefit-to-cost ratio increases with $\bar{A}$, consistent with previous findings. (**c**) Results for a scale-free network, where risk preferences are assigned according to node degree. Each point represents $10^7$ Monte Carlo realizations. "Proportional" indicates that $A$ is proportional to node degree, whereas "inversely proportional" indicates that $A$ is inversely proportional to node degree. High-degree nodes exert a stronger influence on cooperation through their risk preferences. (**d**) Results for a random regular network, where risk preferences are assigned according to node closeness centrality. Each point represents $10^7$ Monte Carlo realizations. Definitions of "proportional" and "inversely proportional" are similar as above. For nodes with identical degrees, lower closeness centrality leads to risk preferences with a stronger impact on cooperation. Parameters: $B = 0.2$, $p = 0.5$, and $Q = 0$ in **b** and **c** and **d**.

14

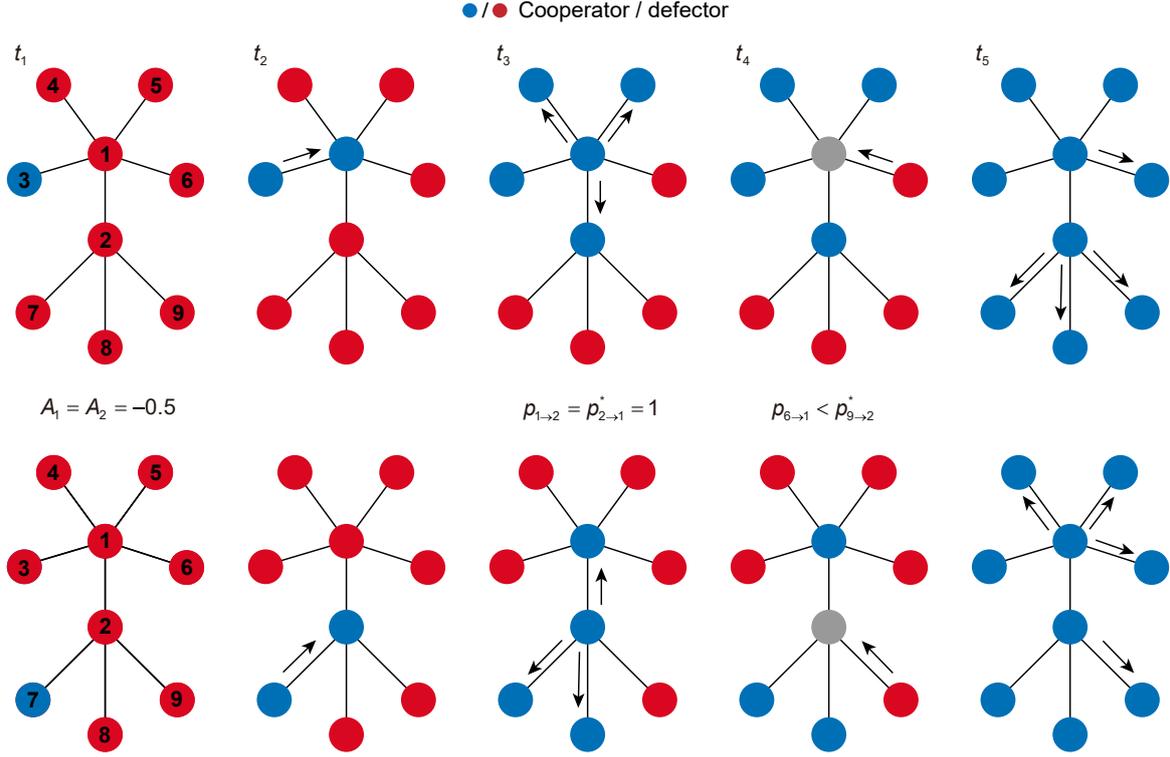

**Supplementary Figure 2: Intuition for the impact of node's degree.** We consider a network with two central nodes (nodes 1 and 2) as shown, with the ∗ marking the lower case. At $t_1$, cooperative mutations occur at nodes 3 and 7, which are transmitted to central nodes 1 and 2 at $t_2$. At $t_3$, central nodes first transmit the strategy to directly connected leaf nodes, then to the other central node. Since the strategy utility of the central nodes is positive in both cases, $p_{1 \to 2} = p^*_{2 \to 1}$, while higher-degree central nodes can transmit the strategy to more nodes. At $t_4$, when central nodes are chosen to update their strategies, it is more difficult for the central node in the upper scenario to adopt the defection strategy, i.e., $p_{6 \to 1} = p^*_{9 \to 2}$. Thus, high-degree nodes are more likely to maintain cooperative strategies when facing defecting neighbors. Parameters: $b = 10$, $c = 1$, $B = 0.5$, $p = 0.5$, $Q = 0$, $A_1 = A_2 = -0.5$.

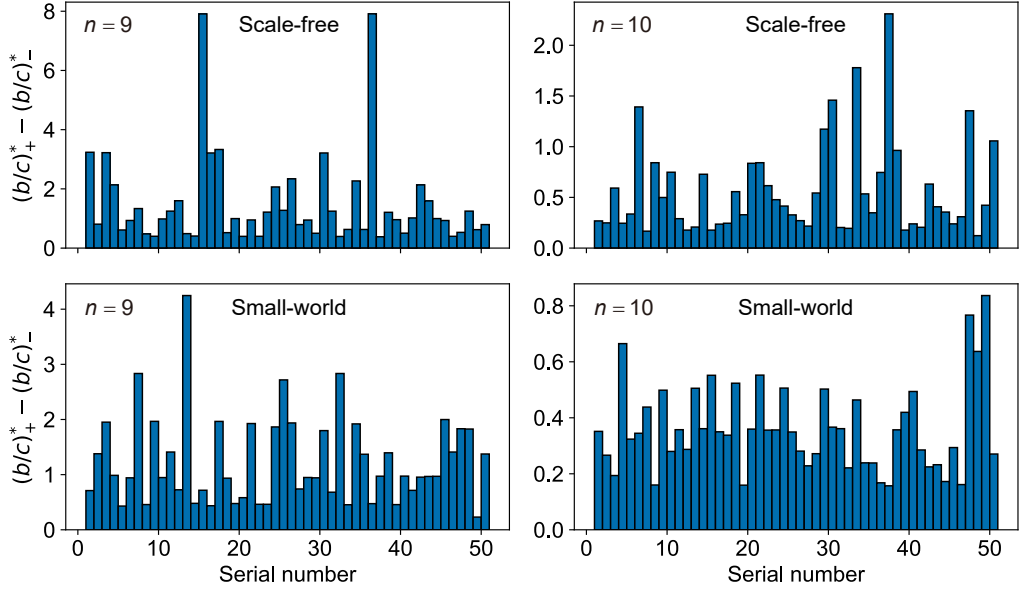

**Supplementary Figure 3: High-degree nodes' risk preferences has stronger influence on cooperation in two types of random networks.** We generated 50 random scale-free and small-world networks with sizes $n = 9$ and $n = 10$. For each graph, risk preferences are assigned either positively or negatively proportional to node degrees, but the average risk preference is kept neutral. Results show that the critical ratios in the positively related case $((b/c)^*_+)$ values exceed those in the negatively related case $((b/c)^*_-)$, indicating that the promotion of cooperation in the population strongly depends on the risk preferences of high-degree nodes. Parameters: $B = 0.2$, $p = 0.5$, and $Q = 0$

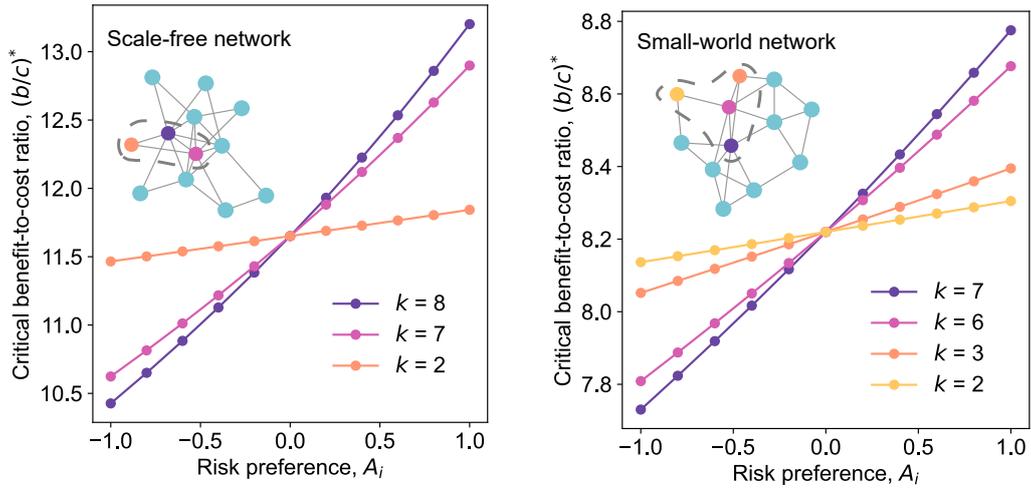

**Supplementary Figure 4: High-degree nodes' risk preferences has stronger influence on cooperation.** In the scale-free and small-world networks shown above, the risk preference of one highlighted node was varied uniformly from $-1$ to $1$ in increments of $0.2$, while keeping all other nodes' risk preference fixed at zero. The resulting two sets of curves demonstrate that nodes with higher degrees exhibit steeper slopes, indicating a stronger influence of their risk preference on the population's critical benefit-to-cost ratio. Parameters: $Q = 0$, $p = 0.5$ and $B = 0.5$.


\end{document}